\documentclass[twocolumn,showpacs,prl]{revtex4}

\ifx\pdfoutput\undefined
\usepackage{graphicx}
\else
\usepackage{graphicx}
\fi

\usepackage[center]{subfigure}
\usepackage{bm}
\usepackage{amsmath,amssymb}
\usepackage{latexsym}

\def\beq{\begin{equation}}
\def\eeq{\end{equation}}
\def\nn{\nonumber}

\begin{document}

\title{Robust ecological pattern formation induced by demographic noise}
\author{Thomas Butler and Nigel Goldenfeld}
\affiliation{Department of Physics and Institute for Genomic Biology,
University of Illinois at Urbana Champaign, 1110 West Green Street, Urbana, IL 61801 USA}

\date{\today}

\begin{abstract}

We demonstrate that demographic noise can induce persistent spatial
pattern formation and temporal oscillations in the Levin-Segel
predator-prey model for plankton-herbivore population dynamics.
Although the model exhibits a Turing instability in mean field theory,
demographic noise greatly enlarges the region of parameter space where
pattern formation occurs. To distinguish between patterns generated by
fluctuations and those present at the mean field level in real
ecosystems, we calculate the power spectrum in the noise-driven case
and predict the presence of fat tails not present in the mean field
case.  These results may account for the prevalence of large-scale
ecological patterns, beyond that expected from traditional
non-stochastic approaches.
\end{abstract}



\maketitle

Many years ago, Turing showed how diffusion, normally thought of as a
homogenizing influence, can give rise to pattern-forming
instabilities\cite{TURI53}.  Only recently, however, have field
observations provided strong support for the presence of Turing
patterns in ecosystems, where diffusional processes
abound, at least in principle.  The slow moving tussock moth population
in California together with its faster moving parasites \cite{MARO97}
as well as several plant-resource systems \cite{REIT08} have been
identified as satisfying, qualitatively at least, the key requirements
for diffusion driven pattern formation. Observed patterns of plankton
populations have also been proposed to arise from Turing instabilities,
at least over short length scales \cite{LEVI76,MALC08,DAVI92,ABRA98}.

The common feature of these systems is positive feedback coupled to
slow diffusion (usually associated with a species labeled an
``activator" that activates both itself and another species called the
``inhibitor"), and negative feedback coupled to faster diffusion
associated with the inhibitor.  This combination of diffusion and
feedback promotes the formation of patterns, because local patches are
promoted through positive feedback, but are only able to spread a
limited distance before the fast diffusion and associated negative
feedback of the inhibitor prevents further spread.  It is hypothesized
that this mechanism is responsible for a great deal of ecosystem
level pattern formation \cite{REIT08,MARO97}.

One particular class of ecological pattern forming systems,
predator-prey (or organism-natural enemy) systems has been extensively
analyzed theoretically (see for example, \cite{LEVI76,MIMU78, BAUR07,
MALC08, MOBI06}) and is beginning to allow qualitative comparison to
field data along with more system specific theory \cite{MARO97,WILS99}.
A difficulty in directly comparing the results of this large body of
theory to field observations is that in many cases, models only exhibit
Turing instabilities if the predator diffusivity is much larger than
the prey diffusivity or the parameters are fine tuned
\cite{LEVI76,MIMU78,WILS99, BAUR07}.  The qualitative argument
made above for pattern formation does not depend on very
large differences in diffusivities, nor on additional ecological
details, and indeed, there are ecological
pattern-forming systems which do not apparently display very large
separation of diffusivities \cite{MARO97,REIT08}.  So what is the
origin of pattern formation in such systems?

One approach to such questions of ordering is to include
levels of detail that in some sense force the response of the system.
For example, whereas simple mean field predator-prey models do not show
population oscillations, they can be made to do so by the inclusion of
predator satiation effects\cite{maynard1974models}.  However, such
levels of realism do not need to be invoked, because there is a simpler
explanation: intrinsic or demographic noise.  This may seem counterintuitive,
because adding noise to a system is usually thought of as reducing
ordering by adding entropy; and indeed, this is exactly what is
observed in several models, such as percolation models of epidemics
\cite{DAVI08} and spin models of forest canopy gaps \cite{KATO98}.
Surprisingly, however, systematic treatments of individual-level models
(ILMs) of predator-prey dynamics show that the population fluctuations
become amplified\cite{MCKAN05}, and lead to time-dependent oscillations
(quasi-cycles) that can be distinguished from deterministic limit cycle
behavior\cite{PINE07}.  Disappointingly, to date, no novel spatial
effects of demographic noise have been identified, despite several
attempts\cite{LUGO08,BUTL09}.

In this Rapid Communication, we demonstrate that noise-induced pattern
formation arises in a simple but biologically-relevant predator-prey model,
and show that if it is analyzed as an ILM, patterns occur over a much
larger range of ecologically relevant parameters than predicted by MFT,
even in the thermodynamic limit.  We accomplish this by calculating the
phase diagram and power spectrum of the model analytically.  We also
predict that experimental noise driven patterns will have power spectra
with fat tails not present in patterns driven by instabilities present
in MFT. Finally, we show that quasi-cycles are also present, and we provide
an interpretation of the spatiotemporal dynamics that result.

\section{Heuristic analysis of the Levin-Segel model}

Among the simplest models of ecological pattern formation was
originally introduced to model plankton-herbivore
dynamics\cite{LEVI76}.  This model takes the form

\begin{equation}
\begin{split}
\partial_t\psi &=\mu\nabla^2\psi+b_1\psi+e\psi^2-(p_1+p_2)\psi\varphi \\
\partial_t\varphi &= \nu\nabla^2\varphi+p_2\varphi\psi-d\varphi^2
\end{split}
\label{1}
\end{equation}

\noindent where the plankton population density is given by $\psi$, the
herbivore population density is given by $\varphi$, $b_1$ is birthrate
for the plankton, $p_1$ and $p_2$ are predation, $d$ is
competition-driven death of the predators and $e$ corresponds to a
community effect, that is the prey facilitates its own birth rate.  In
the original presentation of this model, this term was intended to be a
proxy for reduced predator efficiency at higher prey concentrations
\cite{LEVI76}.  It can also be interpreted as an Allee effect, wherein
many species have enhanced reproduction at higher concentrations (for a
review, see \cite{COUR99}).  From here on, we set $p_1=0$ and $p_2=p$
for transparency of analysis.  This does not change the qualitative
results.  The parameters $e$ and $p, \  d$ identify the prey as the
activator and the predator as the inhibitor in the mechanism for
pattern formation above and distinguish this model from the standard
Lotka-Volterra based individual level models recently analyzed and
demonstrated not to contain patterns in \cite{LUGO08,BUTL09}.

The model contains a stable homogeneous coexistence state when

\begin{equation}
\begin{split}
p>e \ \text{and} \ p^2>de
\end{split}
\label{1a}
\end{equation}

\noindent with fixed point populations given by

\begin{eqnarray}
\psi=\frac{b_1 d}{p^2-de}, \ \ \varphi=\frac{b_1 p}{p^2-de}
\label{1b}
\end{eqnarray}

\noindent It contains a Turing instability if \cite{LEVI76}

\begin{eqnarray}
\frac{\nu}{\mu}>\left(\frac{1}{\left(\sqrt{p/d}-\sqrt{p/d-e/p}\right)}\right)^2
\label{1c}
\end{eqnarray}

When the model violates the stability conditions in Eq. \ref{1a}, the plankton population diverges and a plankton regulation term (i.e. $-f\psi^3$) is required to make the model valid.  Such a term would only materially affect the outcomes of this analysis near the instability, where it would decrease the set of parameters for which pattern formation occurs.  To examine
the behavior of the model, we take the generic set of $O(1)$ kinetic
parameters $b_1=1/2, \ e=1/2, \ d=1/2$ and $p=1$.  With these generic
parameters Eq. \ref{1c} shows that non-generic diffusivities,
$\nu/\mu>27.8$, are required for pattern formation.  Similar results
are obtained for other stable, generic parameter sets.

Demographic noise may change this
picture\cite{WILS1998} by inhibiting the decay of transient patterns.
Turing instabilities occur when, for some specific set of wave vectors,
small perturbations no longer decay.  However, we expect that even when
the parameters are tuned away from the Turing instability,
perturbations with some wavelengths may decay more slowly than others,
leading to transient patterns.  Demographic noise would maintain these
patterns by generating continual perturbations.  This is reminiscent of
extrinsic noise driven patterns reported in other contexts
\cite{GARC93,CARR04,SEIB07}.

To quantify this heuristic argument, we look at the
Fourier transformed dynamics of the fluctuations from the coexistence
fixed point with added white noise $\xi$, variance 1.  These dynamics
are given by

\begin{equation}
-i\omega \bf{x}=\bf{A}\bf{x}+\xi
\label{1d}
\end{equation}

\noindent The matrix $\bf{A}$ is the Fourier transformed stability
matrix

\begin{equation}
\bf{A} = \left( \begin{array}{cc} -\nu k^2 -p\psi& p \varphi \\
                                                                    -p\psi               & -\mu k^2+e\psi
                                                                    \end{array} \right)
\label{1e}
\end{equation}

\noindent Simple manipulations yield the power spectrum

\begin{equation}
\begin{split}
&P(k,\omega)= \left[p^2 \varphi^2+(e\psi-\mu k^2)^2 \right] \times \bigg[\big(p b_1\psi+\mu\nu k^4-\omega^2 \\ &-\psi k^2 e\nu\left(1-\frac{p\mu}{e\nu}\right)\big)^2 +\omega^2((e-p)\psi-(\mu+\nu)k^2)\bigg]^{-2}
\end{split}
\label{1f}
\end{equation}

Very approximately, we expect from Eq. \ref{1f} that patterns
(indicated by peaks in the power spectrum) form whenever $e\nu>p\mu$.
This is much less stringent than Eq. \ref{1c} and can be satisfied for
generic sets of parameters. However, to reliably demonstrate our
hypotheses and extract experimental predictions, we next perform a
systematic study of demographic noise from an individual level model.

\section{Individual Level Model}

We define the individual level version of the model by considering a
locally well mixed patch of volume $V$.  We consider the following
reactions

\begin{align}
P  &\stackrel{b_1}{\rightarrow}PP \ \ &PP &\stackrel{e/V}{\rightarrow} PPP \nn \\
PH &\stackrel{p/V}{\rightarrow}HH \ \ &HH &\stackrel{d/V}{\rightarrow} H
\label{2}
\end{align}

\noindent where $P$ denotes plankton and $H$ denotes herbivores, with
the parameters as described above. Stochastic trajectories of $H$ and
$P$, enumerated by $m$ and $n$ respectively, are described by the
master equation

\begin{align}
&\partial_t P(m,n) = b_1(-nP(m,n)+(n-1)P(m,n-1)) \nn \\
&+\frac{e}{V}[(n-1)(n-2)P(m,n-1) - n(n-1)P(m,n)] \nn \\
&+\frac{p}{V}(-mnP(m,n)+(m-1)(n+1)P(m-1,n+1)) \nn \\
&+\frac{d}{V}\left[(m+1)mP(m+1,n)-m(m-1)P(m,n)\right]
\label{3}
\end{align}

To analyze the master equation, we map it to a
path integral formulation of bosonic field theory  and generalize to space \cite{DOI76, GOLD84, MIKH81, PELI85, JANS08}.  To add space we consider a lattice of patches, and random hopping for both
species at different rates between nearest neighbor patches.  The
resulting Lagrangian density is given by

\begin{align}
\mathcal{L} = &\hat{x}\partial_t z + \hat{\rho} \partial_t \rho -\nu\hat{z}\nabla^2 z - \mu \hat{\rho}\nabla^2 \rho -\nu z(\nabla \hat{z})^2 \nn \\ &-\mu\rho(\nabla \hat{\rho})^2+b_1\rho(1-e^{\hat{\rho}})+\frac{e}{V}\rho^2(1-e^{\hat{\rho}}) \nn \\ &+\frac{p}{V}z\rho(1-e^{\hat{z}-\hat{\rho}})+\frac{d}{V}z^2(1-e^{-\hat{z}})
\label{10}
\end{align}

Where $\hat{z},\; z$ are noise and number variables respectively for herbivores, and similarly, $\hat{\rho}, \; \rho$ are noise and number variables for plankton.  To analyze this Lagrangian directly is difficult,
due to exponential terms and diffusive noise.  To make progress, we
derive a systematic expansion and mean field theory (MFT) in powers of
$\sqrt{V}$ motivated by the $\Omega$-expansion \cite{VANK92,BUTL09}.  We
assume the forms

\begin{align}
\hat{z}& \rightarrow \frac{\hat{z}}{\sqrt{V}}
&\hat{\rho} &\rightarrow \frac{\hat{\rho}}{\sqrt{V}}  \\
z&=V\varphi +\sqrt{V}\eta
&\rho &= V \psi + \sqrt{V}\xi
\label{11}
\end{align}

\noindent for the fields and drop terms with negative powers of
$\sqrt{V}$.  This yields the following form of the Lagrangian

\begin{equation}
\mathcal{L}=\sqrt{V}\mathcal{L}_1+\mathcal{L}_2 +O(1/\sqrt{V})
\label{12}
\end{equation}

\noindent Minimizing $\mathcal{L}_1$ in the infinite $V$ limit yields
the MFT in Eqs. \ref{1}.  Since we've already analyzed it, we now turn
to $\mathcal{L}_2$.  We represent it in matrix form as

\begin{equation}
\mathcal{L}_2 = \bf{y}^T\partial_t \bf{x} - \bf{y}^T\bf{A}\bf{x}-\frac{1}{2}\bf{y}^T\bf{B}\bf{y}
\label{14}
\end{equation}

\noindent The matrix $\bf{A}$ is the stability matrix we used in the
heuristic analysis above, Eq. \ref{1e}.  The matrix $\bf{B}$ is given
by

\begin{equation}
\bf{B}=\left( \begin{array}{cc} 2p\varphi\psi + \nu\varphi k^2& -p \varphi\psi \\
                                                                    -p \varphi\psi               & 2p\varphi\psi +\mu\psi k^2
                                                                    \end{array} \right)
\label{17}
\end{equation}

\noindent where we have Fourier transformed the equations.  We also now
note that $\mathcal{L}_2$ is in the form of a Lagrangian in the
Martin-Siggia-Rose (MSR) response function formalism for Langevin
equations \cite{MART73, JANS76}.  Thus we can extract coupled Langevin
equations for the fluctuations from the Lagrangian by applying the MSR
formalism. The resulting Langevin equations with the appropriate noise
and correlations are

\begin{align}
&-i\omega\bf{x}=\bf{A}\bf{x}+\bf{\gamma(\omega)} \nn \\
&\langle \gamma_i (\omega) \gamma_j(-\omega)\rangle = B_{ij}
\label{18}
\end{align}

\noindent Simple manipulations yield the power spectrum

\begin{equation}
\langle x_1 x_1^* \rangle
= \frac{|D_{22}|^2 B_{11}-2D_{12}Re(D_{22})B_{21}+|D_{12}|^2 B_{22}}{|det(D)|^2}
\label{20}
\end{equation}

This expression results in a rational polynomial with complicated coefficients that
is sixth order in $k$ in the numerator, and eighth in the denominator.   The
denominator is the same as the denominator for the heuristic power spectrum in
Eq. \ref{1f}.  Alternatively, these results could have been obtained by a standard $\Omega$ expansion\cite{VANK92} of the master equation \ref{3}.

\section{Discussion}

Pattern formation occurs when there is a peak in $P(k,\omega)$ at non-zero $k$.  This
occurs if $dP/dk^2>0$ at $k=0$, because for large $k$, the power spectrum
is a decreasing function and has a negative derivative.  The peak occurs at the point
where the derivative changes sign.  Carrying out the derivative at $k=0$ yields

\begin{eqnarray}
\frac{\nu}{\mu}>\frac{p^3(5p^2 + 7de)}{e(4p^4 + 5p^2 de + 3 d^2 e^2)}
\label{23}
\end{eqnarray}

\begin{figure}[ht]
\begin{center}
\includegraphics[width=2.5in]{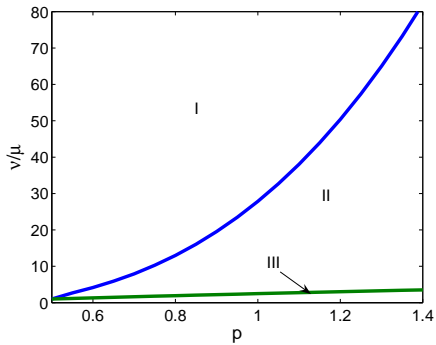}
\caption{Phase diagram over stable parameter region in $p$.  The region I phase is MFT level pattern formation, the region II phase is noise driven pattern formation and quasi-cycles and region III is a spatially homogeneous phase with quasi-cycles.}
\label{fig1}
\end{center}
\end{figure}

Eqs. \ref{23}, \ref{1c} and the stability conditions define the phase
diagram of the model (fig. \ref{fig1}).  For the purposes of the phase
diagram, we fix the parameters as above, leaving $p$ and $\nu/\mu$ as
control parameters.  The phase diagram shows that the beyond mean field
corrections expand the range of ecologically interesting parameters in
which pattern formation occurs greatly.

For larger values of $k$, since the denominator in Eq. \ref{20} goes as
the eighth power, and the numerator as the sixth power of $k$, it is
clear that
\begin{equation}
P\propto k^{-2}
\label{25}
\end{equation}

This provides an experimental prediction: in regions II and III of the phase diagram, the power
spectrum will have a fat tail that decays as approximately $k^{-2}$.  In region I, the power spectrum will be dominated by the spatially structured mean field populations, and should fall off much more quickly.  This is analogous to the statistical test to
distinguish quasi-cycles from limit cycles in predator-prey populations
that recently showed population oscillations in wolverines to be driven
by finite size fluctuations \cite{PINE07,MCKAN05}.

\begin{figure}[ht]
\begin{center}
\includegraphics[width=2.5in]{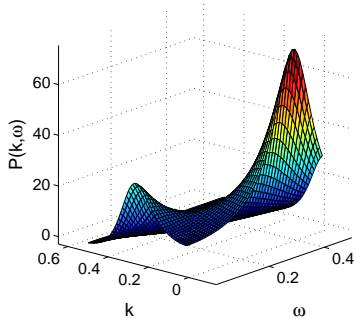}
\caption{Power spectrum with p=1, $\nu/\mu$=15}
\label{fig2}
\end{center}
\end{figure}

An additional feature of the model is that oscillations and spatial
pattern formation are essentially decoupled.  This means that the model
predicts global population oscillations and spatial pattern formation,
but not traveling waves.  The mathematical origin of this can be seen
in Eq. \ref{1f}.  The $k^2$ term with a negative coefficient at
$\omega=0$ is quickly overwhelmed by the positive $k^2$ dependence of
the $\omega^2$ term as the frequency begins to grow.  In the power
spectrum (fig. \ref{fig2}) this can be seen as the deep valley between
the peaks in $k$ and $\omega$.  This interpretation is supported by
preliminary simulations of an agent based model.

We also note that the appropriate thermodynamic limit of the theory is
not $V\rightarrow \infty$, but rather that the number of patches of
size $V$ goes to infinity.  Since $V$ is the volume of a locally well
mixed population, it should never be infinity for a system in which
diffusion effects are significant.  Thus the results we have presented do not depend on the size of the population being studied, and even apply to infinite populations, provided local populations are finite.  In ecological terms, this means
that systems in which fluctuation effects might be expected to be
insignificant due to large populations (e.g. plankton) are equally
likely to contain fluctuation driven patterns and cycles as systems
with small populations, at least over length scales where diffusion is
a reasonable approximation for the spatial dynamics.

The results we have given here were calculated within a specific model,
but we expect that they will be substantially unchanged in any model
with a slow diffusing activator species and a faster diffusing inhibitor
species.



This work was partially supported by National Science Foundation grant NSF-EF-0526747.

\bibliographystyle{apsrev}

\bibliography{PatternPaperBib}

\begin{thebibliography}{31}
\expandafter\ifx\csname natexlab\endcsname\relax\def\natexlab#1{#1}\fi
\expandafter\ifx\csname bibnamefont\endcsname\relax
  \def\bibnamefont#1{#1}\fi
\expandafter\ifx\csname bibfnamefont\endcsname\relax
  \def\bibfnamefont#1{#1}\fi
\expandafter\ifx\csname citenamefont\endcsname\relax
  \def\citenamefont#1{#1}\fi
\expandafter\ifx\csname url\endcsname\relax
  \def\url#1{\texttt{#1}}\fi
\expandafter\ifx\csname urlprefix\endcsname\relax\def\urlprefix{URL }\fi
\providecommand{\bibinfo}[2]{#2}
\providecommand{\eprint}[2][]{\url{#2}}

\bibitem[{\citenamefont{Turing}(1953)}]{TURI53}
\bibinfo{author}{\bibfnamefont{A.~M.} \bibnamefont{Turing}},
  \bibinfo{journal}{Phil. Trans. Roy. Soc. B} \textbf{\bibinfo{volume}{237}},
  \bibinfo{pages}{37} (\bibinfo{year}{1953}).

\bibitem[{\citenamefont{Maron and Harrison}(1997)}]{MARO97}
\bibinfo{author}{\bibfnamefont{J.~L.} \bibnamefont{Maron}} \bibnamefont{and}
  \bibinfo{author}{\bibfnamefont{S.}~\bibnamefont{Harrison}},
  \bibinfo{journal}{Science} \textbf{\bibinfo{volume}{278}},
  \bibinfo{pages}{1619} (\bibinfo{year}{1997}).

\bibitem[{\citenamefont{Reitkerk and van~de Koppel}(2008)}]{REIT08}
\bibinfo{author}{\bibfnamefont{M.}~\bibnamefont{Reitkerk}} \bibnamefont{and}
  \bibinfo{author}{\bibfnamefont{J.}~\bibnamefont{van~de Koppel}},
  \bibinfo{journal}{TREE} \textbf{\bibinfo{volume}{23}}, \bibinfo{pages}{169}
  (\bibinfo{year}{2008}).

\bibitem[{\citenamefont{Levin and Segel}(1976)}]{LEVI76}
\bibinfo{author}{\bibfnamefont{S.~A.} \bibnamefont{Levin}} \bibnamefont{and}
  \bibinfo{author}{\bibfnamefont{L.~A.} \bibnamefont{Segel}},
  \bibinfo{journal}{Nature} \textbf{\bibinfo{volume}{259}},
  \bibinfo{pages}{659} (\bibinfo{year}{1976}).

\bibitem[{\citenamefont{Malchow et~al.}(1998)\citenamefont{Malchow, Hilker,
  Siekmann, Petrovski, and Medvinsky}}]{MALC08}
\bibinfo{author}{\bibfnamefont{H.}~\bibnamefont{Malchow}},
  \bibinfo{author}{\bibfnamefont{F.~M.} \bibnamefont{Hilker}},
  \bibinfo{author}{\bibfnamefont{I.}~\bibnamefont{Siekmann}},
  \bibinfo{author}{\bibfnamefont{S.}~\bibnamefont{Petrovski}},
  \bibnamefont{and} \bibinfo{author}{\bibfnamefont{A.~B.}
  \bibnamefont{Medvinsky}}, \bibinfo{journal}{Aspects of Mathematical
  Modelling} pp. \bibinfo{pages}{1--26} (\bibinfo{year}{1998}).

\bibitem[{\citenamefont{Davis et~al.}(1992)\citenamefont{Davis, Gallager, and
  Solow}}]{DAVI92}
\bibinfo{author}{\bibfnamefont{C.~S.} \bibnamefont{Davis}},
  \bibinfo{author}{\bibfnamefont{S.~M.} \bibnamefont{Gallager}},
  \bibnamefont{and} \bibinfo{author}{\bibfnamefont{A.~R.} \bibnamefont{Solow}},
  \bibinfo{journal}{Science} \textbf{\bibinfo{volume}{257}},
  \bibinfo{pages}{230} (\bibinfo{year}{1992}).

\bibitem[{\citenamefont{Abraham}(1998)}]{ABRA98}
\bibinfo{author}{\bibfnamefont{E.~R.} \bibnamefont{Abraham}},
  \bibinfo{journal}{Nature} \textbf{\bibinfo{volume}{391}},
  \bibinfo{pages}{577} (\bibinfo{year}{1998}).

\bibitem[{\citenamefont{Mimura and Murray}(1978)}]{MIMU78}
\bibinfo{author}{\bibfnamefont{M.}~\bibnamefont{Mimura}} \bibnamefont{and}
  \bibinfo{author}{\bibfnamefont{J.~D.} \bibnamefont{Murray}},
  \bibinfo{journal}{J. Theor. Biol.} \textbf{\bibinfo{volume}{75}},
  \bibinfo{pages}{249} (\bibinfo{year}{1978}).

\bibitem[{\citenamefont{Baurmann et~al.}(2007)\citenamefont{Baurmann, Gross,
  and Feudel}}]{BAUR07}
\bibinfo{author}{\bibfnamefont{M.}~\bibnamefont{Baurmann}},
  \bibinfo{author}{\bibfnamefont{T.}~\bibnamefont{Gross}}, \bibnamefont{and}
  \bibinfo{author}{\bibfnamefont{U.}~\bibnamefont{Feudel}},
  \bibinfo{journal}{J. Theor. Biol.} \textbf{\bibinfo{volume}{245}},
  \bibinfo{pages}{220} (\bibinfo{year}{2007}).

\bibitem[{\citenamefont{Mobilia et~al.}(2006)\citenamefont{Mobilia, Georgiev,
  and Tauber}}]{MOBI06}
\bibinfo{author}{\bibfnamefont{M.}~\bibnamefont{Mobilia}},
  \bibinfo{author}{\bibfnamefont{I.~T.} \bibnamefont{Georgiev}},
  \bibnamefont{and} \bibinfo{author}{\bibfnamefont{U.~C.}
  \bibnamefont{Tauber}}, \bibinfo{journal}{Phys.\ Rev.~E}
  \textbf{\bibinfo{volume}{73}}, \bibinfo{pages}{040903(R)}
  (\bibinfo{year}{2006}).

\bibitem[{\citenamefont{Wilson et~al.}(1999)\citenamefont{Wilson, Harrison,
  Hastings, and McCann}}]{WILS99}
\bibinfo{author}{\bibfnamefont{W.~G.} \bibnamefont{Wilson}},
  \bibinfo{author}{\bibfnamefont{S.~P.} \bibnamefont{Harrison}},
  \bibinfo{author}{\bibfnamefont{A.}~\bibnamefont{Hastings}}, \bibnamefont{and}
  \bibinfo{author}{\bibfnamefont{K.}~\bibnamefont{McCann}},
  \bibinfo{journal}{J. Anim. Ecol.} pp. \bibinfo{pages}{94--107}
  (\bibinfo{year}{1999}).

\bibitem[{\citenamefont{Maynard~Smith}(1974)}]{maynard1974models}
\bibinfo{author}{\bibfnamefont{J.}~\bibnamefont{Maynard~Smith}}
  (\bibinfo{year}{1974}).

\bibitem[{\citenamefont{Davis et~al.}(2008)\citenamefont{Davis, Trapman, Leirs,
  Begon, and Heesterbeek}}]{DAVI08}
\bibinfo{author}{\bibfnamefont{S.}~\bibnamefont{Davis}},
  \bibinfo{author}{\bibfnamefont{P.}~\bibnamefont{Trapman}},
  \bibinfo{author}{\bibfnamefont{H.}~\bibnamefont{Leirs}},
  \bibinfo{author}{\bibfnamefont{M.}~\bibnamefont{Begon}}, \bibnamefont{and}
  \bibinfo{author}{\bibfnamefont{J.~A.~P.} \bibnamefont{Heesterbeek}},
  \bibinfo{journal}{Nature} \textbf{\bibinfo{volume}{454}},
  \bibinfo{pages}{634} (\bibinfo{year}{2008}).

\bibitem[{\citenamefont{Katori et~al.}(1998)\citenamefont{Katori, Kizaki,
  Terui, and Kubo}}]{KATO98}
\bibinfo{author}{\bibfnamefont{M.}~\bibnamefont{Katori}},
  \bibinfo{author}{\bibfnamefont{S.}~\bibnamefont{Kizaki}},
  \bibinfo{author}{\bibfnamefont{Y.}~\bibnamefont{Terui}}, \bibnamefont{and}
  \bibinfo{author}{\bibfnamefont{T.}~\bibnamefont{Kubo}},
  \bibinfo{journal}{Fractals} \textbf{\bibinfo{volume}{6}}, \bibinfo{pages}{81}
  (\bibinfo{year}{1998}).

\bibitem[{\citenamefont{McKane and Newman}(2005)}]{MCKAN05}
\bibinfo{author}{\bibfnamefont{A.~J.} \bibnamefont{McKane}} \bibnamefont{and}
  \bibinfo{author}{\bibfnamefont{T.~J.} \bibnamefont{Newman}},
  \bibinfo{journal}{Phys.\ Rev.\ Lett.} \textbf{\bibinfo{volume}{94}},
  \bibinfo{pages}{218102} (\bibinfo{year}{2005}).

\bibitem[{\citenamefont{Pineda-Krch et~al.}(2007)\citenamefont{Pineda-Krch,
  Blok, and Doebeli}}]{PINE07}
\bibinfo{author}{\bibfnamefont{M.}~\bibnamefont{Pineda-Krch}},
  \bibinfo{author}{\bibfnamefont{H.~J.} \bibnamefont{Blok}}, \bibnamefont{and}
  \bibinfo{author}{\bibfnamefont{M.}~\bibnamefont{Doebeli}},
  \bibinfo{journal}{Oikos} \textbf{\bibinfo{volume}{116}}, \bibinfo{pages}{53}
  (\bibinfo{year}{2007}).

\bibitem[{\citenamefont{Lugo and McKane}(2008)}]{LUGO08}
\bibinfo{author}{\bibfnamefont{C.}~\bibnamefont{Lugo}} \bibnamefont{and}
  \bibinfo{author}{\bibfnamefont{A.~J.} \bibnamefont{McKane}},
  \bibinfo{journal}{Phys.\ Rev.~E} \textbf{\bibinfo{volume}{78}}
  (\bibinfo{year}{2008}).

\bibitem[{\citenamefont{Butler and Reynolds}(2009)}]{BUTL09}
\bibinfo{author}{\bibfnamefont{T.}~\bibnamefont{Butler}} \bibnamefont{and}
  \bibinfo{author}{\bibfnamefont{D.}~\bibnamefont{Reynolds}},
  \bibinfo{journal}{Phys. Rev. E} \textbf{\bibinfo{volume}{79}},
  \bibinfo{pages}{032901} (\bibinfo{year}{2009}).

\bibitem[{\citenamefont{Courchamp et~al.}(1999)\citenamefont{Courchamp,
  Clutton-Brock, and Grenfell}}]{COUR99}
\bibinfo{author}{\bibfnamefont{F.}~\bibnamefont{Courchamp}},
  \bibinfo{author}{\bibfnamefont{T.}~\bibnamefont{Clutton-Brock}},
  \bibnamefont{and} \bibinfo{author}{\bibfnamefont{B.}~\bibnamefont{Grenfell}},
  \bibinfo{journal}{TREE} \textbf{\bibinfo{volume}{14}}, \bibinfo{pages}{405}
  (\bibinfo{year}{1999}).

\bibitem[{\citenamefont{Wilson}(1998)}]{WILS1998}
\bibinfo{author}{\bibfnamefont{W.~G.} \bibnamefont{Wilson}},
  \bibinfo{journal}{The American Naturalist} \textbf{\bibinfo{volume}{151}},
  \bibinfo{pages}{116} (\bibinfo{year}{1998}).

\bibitem[{\citenamefont{Garc\'ia-Ojalvo
  et~al.}(1993)\citenamefont{Garc\'ia-Ojalvo, Hern\'andez-Machado, and
  Sancho}}]{GARC93}
\bibinfo{author}{\bibfnamefont{J.}~\bibnamefont{Garc\'ia-Ojalvo}},
  \bibinfo{author}{\bibfnamefont{A.}~\bibnamefont{Hern\'andez-Machado}},
  \bibnamefont{and} \bibinfo{author}{\bibfnamefont{J.~M.}
  \bibnamefont{Sancho}}, \bibinfo{journal}{Phys. Rev. Lett.}
  \textbf{\bibinfo{volume}{71}}, \bibinfo{pages}{1542} (\bibinfo{year}{1993}).

\bibitem[{\citenamefont{Carrillo et~al.}(2004)\citenamefont{Carrillo, A, J, and
  Sancho}}]{CARR04}
\bibinfo{author}{\bibfnamefont{O.}~\bibnamefont{Carrillo}},
  \bibinfo{author}{\bibfnamefont{S.~M.} \bibnamefont{A}},
  \bibinfo{author}{\bibfnamefont{G.-O.} \bibnamefont{J}}, \bibnamefont{and}
  \bibinfo{author}{\bibfnamefont{J.~M.} \bibnamefont{Sancho}},
  \bibinfo{journal}{Europhys. Lett.} \textbf{\bibinfo{volume}{65}},
  \bibinfo{pages}{452} (\bibinfo{year}{2004}).

\bibitem[{\citenamefont{Sieber et~al.}(2007)\citenamefont{Sieber, Malchow, and
  Schimansky-Geier}}]{SEIB07}
\bibinfo{author}{\bibfnamefont{M.}~\bibnamefont{Sieber}},
  \bibinfo{author}{\bibfnamefont{H.}~\bibnamefont{Malchow}}, \bibnamefont{and}
  \bibinfo{author}{\bibfnamefont{L.}~\bibnamefont{Schimansky-Geier}},
  \bibinfo{journal}{Ecological complexity} \textbf{\bibinfo{volume}{4}},
  \bibinfo{pages}{223} (\bibinfo{year}{2007}).

\bibitem[{\citenamefont{Doi}(1976)}]{DOI76}
\bibinfo{author}{\bibfnamefont{M.}~\bibnamefont{Doi}}, \bibinfo{journal}{J.
  Phys. A.} \textbf{\bibinfo{volume}{9}}, \bibinfo{pages}{1465}
  (\bibinfo{year}{1976}).

\bibitem[{\citenamefont{Goldenfeld}(1984)}]{GOLD84}
\bibinfo{author}{\bibfnamefont{N.}~\bibnamefont{Goldenfeld}},
  \bibinfo{journal}{J. Phys. A} \textbf{\bibinfo{volume}{17}},
  \bibinfo{pages}{2807} (\bibinfo{year}{1984}).

\bibitem[{\citenamefont{Mikhailov}(1981)}]{MIKH81}
\bibinfo{author}{\bibfnamefont{A.~S.} \bibnamefont{Mikhailov}},
  \bibinfo{journal}{Phys. Lett.} \textbf{\bibinfo{volume}{85}},
  \bibinfo{pages}{214} (\bibinfo{year}{1981}).

\bibitem[{\citenamefont{Peliti}(1985)}]{PELI85}
\bibinfo{author}{\bibfnamefont{L.}~\bibnamefont{Peliti}}, \bibinfo{journal}{PJ.
  Physique} \textbf{\bibinfo{volume}{46}}, \bibinfo{pages}{1469}
  (\bibinfo{year}{1985}).

\bibitem[{\citenamefont{Janssen and Tauber}(2005)}]{JANS08}
\bibinfo{author}{\bibfnamefont{H.~K.} \bibnamefont{Janssen}} \bibnamefont{and}
  \bibinfo{author}{\bibfnamefont{U.~C.} \bibnamefont{Tauber}},
  \bibinfo{journal}{Annals of Physics} \textbf{\bibinfo{volume}{315}},
  \bibinfo{pages}{147} (\bibinfo{year}{2005}).

\bibitem[{\citenamefont{Van~Kampen}(1992)}]{VANK92}
\bibinfo{author}{\bibfnamefont{N.~G.} \bibnamefont{Van~Kampen}},
  \emph{\bibinfo{title}{Stochastic Processes in Physics and Chemistry}}
  (\bibinfo{publisher}{Elsevier, New York}, \bibinfo{year}{1992}).

\bibitem[{\citenamefont{Martin et~al.}(1973)\citenamefont{Martin, Siggia, and
  Rose}}]{MART73}
\bibinfo{author}{\bibfnamefont{P.~C.} \bibnamefont{Martin}},
  \bibinfo{author}{\bibfnamefont{E.~D.} \bibnamefont{Siggia}},
  \bibnamefont{and} \bibinfo{author}{\bibfnamefont{H.~A.} \bibnamefont{Rose}},
  \bibinfo{journal}{Phys.\ Rev.\ A} \textbf{\bibinfo{volume}{8}},
  \bibinfo{pages}{423} (\bibinfo{year}{1973}).

\bibitem[{\citenamefont{Bausch et~al.}(1976)\citenamefont{Bausch, Janssen, and
  Wagner}}]{JANS76}
\bibinfo{author}{\bibfnamefont{R.}~\bibnamefont{Bausch}},
  \bibinfo{author}{\bibfnamefont{H.~K.} \bibnamefont{Janssen}},
  \bibnamefont{and} \bibinfo{author}{\bibfnamefont{H.}~\bibnamefont{Wagner}},
  \bibinfo{journal}{Z. Phys. B.} \textbf{\bibinfo{volume}{24}},
  \bibinfo{pages}{113} (\bibinfo{year}{1976}).

\end{thebibliography}

\end{document}